\documentclass[12pt]{article}
\usepackage{bbm,latexsym}
\newcommand{\be}{\begin{equation}}
\newcommand{\ee}{\end{equation}}
\newcommand{\ba}{\begin{eqnarray}}
\newcommand{\ea}{\end{eqnarray}}
\newcommand{\no}{\nonumber\\}
\newcommand{\lesssim}{\:\mbox{\raisebox{-4pt}{$\stackrel%
{\displaystyle <}{\sim}$}}\:}
\newcommand{\gtrsim}{\:\mbox{\raisebox{-4pt}{$\stackrel%
{\displaystyle >}{\sim}$}}\:}
\newcommand{\mbb}{m_{\beta\beta}}
\newcommand{\dsm}{\Delta m^2_\odot}
\newcommand{\dam}{\Delta m^2_\mathrm{atm}}

\begin{document}
\title{\normalsize \hfill UWThPh-2004-35 \\[1cm] \LARGE
On a model with two zeros \\ in the neutrino mass matrix}
\author{Walter Grimus\thanks{E-mail: walter.grimus@univie.ac.at} \\
\setcounter{footnote}{6}
\small Institut f\"ur Theoretische Physik, Universit\"at Wien \\
\small Boltzmanngasse 5, A--1090 Wien, Austria \\*[3.6mm]
Lu\'\i s Lavoura\thanks{E-mail: balio@cftp.ist.utl.pt} \\
\small Universidade T\'ecnica de Lisboa
and Centro de F\'\i sica Te\'orica de Part\'\i culas \\
\small Instituto Superior T\'ecnico,
P--1049-001 Lisboa, Portugal \\*[4.6mm] }

\date{20 April 2005}

\maketitle

\begin{abstract}
We consider a Majorana neutrino mass matrix $\mathcal{M}_\nu$
with $\left( \mathcal{M}_\nu \right)_{\mu\mu} =
\left( \mathcal{M}_\nu \right)_{\tau\tau} = 0$,
in the basis where the charged-lepton mass matrix is diagonal.
We show that this pattern for the lepton mass matrices can be enforced
by extending the Standard Model with three scalar $SU(2)$ triplets
and by using a horizontal symmetry group $\mathbbm{Z}_4$.
The type-II seesaw mechanism
leads to very small vacuum expectation values for the triplets,
thus explaining the smallness of the neutrino masses;
at the same time,
that mechanism renders the physical scalars
originating in the triplets very heavy.
We show that the conditions $\left( \mathcal{M}_\nu \right)_{\mu\mu}
= \left( \mathcal{M}_\nu \right)_{\tau\tau} = 0$
allow both for a normal neutrino mass spectrum and for an inverted one.
In the first case,
the neutrino masses must be larger than $0.1\, \mbox{eV}$
and the atmospheric mixing angle $\theta_{23}$
must be practically equal to $45^\circ$.
In the second case,
the product $\sin{\theta_{13}} \left| \tan{2 \theta_{23}} \right|$ 
must be of order one or larger,
thus correlating the large or maximal atmospheric neutrino mixing
with the smallness of the mixing angle $\theta_{13}$.
\end{abstract}

\newpage
\setcounter{footnote}{4}


\section{Introduction}

Assuming the neutrinos to be Majorana particles, 
the neutrino mass term is given by
\begin{equation}
\mathcal{L}_{\nu\, \mathrm{mass}}
= \frac{1}{2}\, \nu_\mathrm{L}^\mathrm{T} C^{-1}
\mathcal{M}_\nu \nu_\mathrm{L}
+ \mbox{H.c.},
\end{equation}
with a symmetric mass matrix $\mathcal{M}_\nu$.
It has been shown in~\cite{FGM} that,
in the basis where the charged-lepton mass matrix is diagonal,
there are seven possibilities
for a $\mathcal{M}_\nu$ with two `texture' zeros
which are compatible with the present neutrino data.\footnote{See,
for instance,
\cite{maltoni} for a global fit to that data,
and~\cite{grimus} for a general review.}
The phenomenology of those seven mass matrices
has been discussed in~\cite{FGM,xing1,xing2}.
Some of those matrices have various embeddings
in the seesaw mechanism~\cite{seesaw},
by placing zeros in the Majorana mass matrix $M_\mathrm{R}$
of the right-handed neutrino singlets $\nu_\mathrm{R}$
and in the Dirac mass matrix $M_\mathrm{D}$
connecting the $\nu_\mathrm{R}$
with the $\nu_\mathrm{L}$~\cite{honda,GJLT,lavoura}. 

In this letter we concentrate
on the specific neutrino mass matrix called `case C' in~\cite{FGM},
which is
\begin{equation}\label{C}
\mathcal{M}_\nu \sim 
\left( \begin{array}{ccc} 
\times & \times & \times \\ \times & 0 & \times \\ \times & \times & 0 
\end{array} \right),
\end{equation}
where the crosses denote non-zero matrix elements.
The purpose of this letter is twofold.
Firstly,
we present a model,
based on the horizontal symmetry group $\mathbbm{Z}_4$,
which leads to~(\ref{C}).
Secondly,
we provide a comprehensive analytical discussion
of the phenomenological predictions following from~(\ref{C}).

Texture zeros in mass matrices may in general be explained
by Abelian symmetries~\cite{GJLT},
at the expense of an extended scalar sector.
We shall use the second method of~\cite{GJLT}
in order to generate the texture zeros of case C.
That general method prescribes that
a horizontal symmetry group $\mathbbm{Z}_n$
can enforce any desired texture zeros,
provided $n \in \mathbbm{N}$ is sufficiently large.
For the specific case C we have found that $n = 4$ is sufficient;
with $\mathbbm{Z}_4$,
case C can be embedded in a model with a scalar sector consisting of
one single $SU(2)_\mathrm{L}$ doublet---the Standard Model (SM)
Higgs doublet---and three $SU(2)_\mathrm{L}$ triplets.
The use of scalar triplets
for generating models for the neutrino masses and mixings
has been advocated in~\cite{ma};
in that paper,
case C was built into a model based on the non-Abelian group $Q_8$,
with a scalar sector consisting of two doublets and four triplets,
hence much larger than the one of the $\mathbbm{Z}_4$ model presented here.
The merits of scalar triplets are that they can be made very heavy while,
at the same time,
their vacuum expectation values (VEVs) become very small,
through the so-called `type-II seesaw' mechanism~\cite{shafi,ma1}.
The smallness of the triplet VEVs
explains the smallness of the neutrino masses,
as an alternative to the ordinary
(`type-I') seesaw mechanism~\cite{ma1,pfeiffer}.

In section~\ref{model} we present the $\mathbbm{Z}_4$ model for case C.
In section~\ref{phenomenology} we discuss the phenomenology
of the mass matrix~(\ref{C}).
Section~\ref{concl} contains our conclusions.
In an appendix we show that the remaining six viable cases of~\cite{FGM}
can be incorporated in models with scalar triplets,
through the use of appropriate horizontal symmetry groups $\mathbbm{Z}_n$.


\section{The model} \label{model}

The Yukawa couplings of scalar triplets $\Delta_j$---which we write
in a $2 \times 2$ matrix notation---are given by 
\begin{equation}
\mathcal{L}_{\Delta\, \mathrm{Yukawa}} =
\frac{1}{2}\, \sum_j \sum_{\alpha,\beta=e,\mu,\tau}\! 
h_{\alpha\beta}^j\, D_{\alpha \mathrm{L}}^\mathrm{T} C^{-1}
\left( i \tau_2 \Delta_j \right) D_{\beta \mathrm{L}}
+ \mbox{H.c.},
\end{equation}
whence one obtains
\begin{equation}
\mathcal{M}_\nu = \sum_j w_j h^j, 
\quad \mbox{where} \quad
\left\langle 0 \left| \Delta_j \right| 0 \right\rangle
= \left( \begin{array}{cc} 0 & 0 \\ w_j & 0 \end{array} \right).
\end{equation}

Let us assume a horizontal symmetry such that
the lepton doublets $D_{\alpha \mathrm{L}}$
and charged-lepton singlets $\alpha_\mathrm{R}$
($\alpha = e, \mu, \tau$)
transform according to
\begin{equation}\label{trans0}
D_{\alpha \mathrm{L}} \to p_\alpha D_{\alpha \mathrm{L}},
\quad
\alpha_\mathrm{R} \to p_\alpha \alpha_\mathrm{R},
\end{equation}
respectively, 
with $\left| p_\alpha \right| = 1$.
We furthermore assume that the three phase factors $p_e$,
$p_\mu$,
and $p_\tau$ are all different.
This guarantees,
if all Higgs doublets transform trivially under the horizontal symmetry,
that the charged-lepton mass matrix is diagonal.
With the choice
\begin{equation}\label{trans1}
p_e = 1, \quad p_\mu = i, \quad p_\tau = -i,
\end{equation}
the bilinears $D_{\alpha \mathrm{L}}^\mathrm{T}
C^{-1} D_{\beta \mathrm{L}}$ transform as 
\begin{equation}\label{trans2}
\left( \begin{array}{ccc}
1 & i & -i \\ i & -1 & 1 \\ -i & 1 & -1
\end{array} \right).
\end{equation}
Then,
assuming the existence of only three scalar triplets transforming as
\begin{equation}\label{trans3}
\Delta_1 \to \Delta_1, \quad
\Delta_2 \to -i \Delta_2, \quad
\Delta_3 \to i \Delta_3,
\end{equation}
we are able to generate the $\mathcal{M}_\nu$ of case C.
The transformation in~(\ref{trans1}) and~(\ref{trans3})
corresponds to a group $\mathbbm{Z}_4$.
We have thus constructed a model
which incorporates the mass matrix~(\ref{C}).
Besides the three scalar triplets,
the scalar sector of our model
only has the sole Higgs doublet $\phi$ of the SM.

The triplet $\Delta_1$ transforms trivially under $\mathbbm{Z}_4$.
Therefore,
the term $\phi^\dagger \Delta_1 \tilde \phi$
plus its Hermitian conjugate---where $\tilde \phi \equiv
i \tau_2 \phi^\ast$---is allowed in the scalar potential $V$.
Furthermore,
respecting the transformation~(\ref{trans3}),
only the bilinears $\mbox{tr} \left( \Delta_j^\dagger \Delta_k \right)$
with $j = k$ are allowed in the potential.
For reasons to be explained below,
one must break~(\ref{trans3}) softly,
through terms of dimension two,
by allowing all possible bilinears
$\mbox{tr} \left( \Delta_j^\dagger \Delta_k \right)$.
We thus obtain the scalar potential 
\begin{equation}\label{pot}
V =\mu^2 \phi^\dagger \phi + 
\sum_{j,k=1}^3 \left( \mu^2_\Delta \right)_{jk}
\mbox{tr} \left( \Delta_j^\dagger \Delta_k \right) + 
\left( m\, \phi^\dagger \Delta_1 \tilde \phi + \mbox{H.c.} \right)
+ \cdots,
\end{equation}
where the dots indicate the quartic terms which respect $\mathbbm{Z}_4$,
and the $3 \times 3$ matrix $\mu^2_\Delta$ is Hermitian and positive.
Applying the type-II seesaw mechanism~\cite{ma1},
we stipulate that the eigenvalues of $\mu^2_\Delta$,
and all its matrix elements, 
are of an order of magnitude $\mu_T^2$
such that $\mu_T$ is much larger than the electroweak scale,
represented by the VEV $v$ of the lower component of $\phi$
($v \approx 174\, \mbox{GeV}$).
Furthermore,
we require that $\left| m \right| \lesssim\, \mu_T$, whereas 
$\left| \mu^2 \right| \sim v^2$;
all quartic couplings are assumed to be of order one or smaller.
One then obtains~\cite{ma1}
\begin{equation}
w_j \simeq - m^\ast v^2 \left[ \left( \mu^2_\Delta \right)^{-1} \right]_{j1},
\end{equation}
hence $\left| w_j \right| \ll v$.
This shows that in our scenario,
with one trilinear term together with all possible quadratic terms
in the scalar potential,
the type-II seesaw mechanism is operative.
In order that the triplet VEVs are of the order of the neutrino masses,
$0.1\, \mbox{eV}$,
the scale $\mu_T$ must be around $10^{13}\, \mbox{GeV}$~\cite{ma1}.
Such a scale could be an intermediate scale in a Grand Unified Theory
based on $SO(10)$, $E_6$, or some other large group.

Without the soft-breaking quadratic terms in $V$,
i.e.\ without the $\left( \mu^2_\Delta \right)_{jk}$ for $j \neq k$,
both $w_2$ and $w_3$ vanish.
The soft breaking of the horizontal symmetry
is necessary in order to obtain non-zero VEVs $w_2$ and $w_3$
and thus the full mass matrix~(\ref{C}).


\section{Phenomenology of the model} \label{phenomenology}

The mass matrix $\mathcal{M}_\nu$ is diagonalized by 
$U = e^{i \hat \alpha}\, \bar U\,
\mbox{diag} \left( e^{i \rho}, e^{i \sigma}, 1 \right)$,
where $e^{i \hat \alpha}$ is a physically meaningless unitary
diagonal matrix and
\be
\bar U = \left( \begin{array}{ccc}
c_{13} c_{12} &
c_{13} s_{12} &
s_{13} e^{-i \delta} \\
- c_{23} s_{12} - s_{23} s_{13} c_{12} e^{i \delta} &
c_{23} c_{12} - s_{23} s_{13} s_{12} e^{i \delta} &
s_{23} c_{13} \\
-s_{23} s_{12} + c_{23} s_{13} c_{12} e^{i \delta} &
 s_{23} c_{12} + c_{23} s_{13} s_{12} e^{i \delta} &
-c_{23} c_{13} 
\end{array} \right),
\ee
with $c_{ij} \equiv \cos{\theta_{ij}}$
and $s_{ij} \equiv \sin{\theta_{ij}}$, 
the $\theta_{ij}$ being angles of the first quadrant.
The information on the mixing angles $\theta_{ij}$
can be summarized as~\cite{maltoni}
\ba
\theta_{12} &\simeq& 33^\circ \pm 3^\circ\ \mbox{at}\ 90\%\ \mbox{CL},
\no
\theta_{23} &\simeq& 45^\circ \pm 8^\circ\ \mbox{at}\ 90\%\ \mbox{CL},
\label{angles} \\
s_{13}^2 &<& 0.047\ \mbox{at}\ 3\sigma\ \mbox{level}.
\nonumber
\ea
The best-fit values for the mass-squared differences are
\be
\label{m2diff}
\begin{array}{rcccl}
\dam &\equiv& \left| m_3^2 - m_2^2 \right|
&=& 2.2 \times 10^{-3}\, \mbox{eV}^2, 
\\*[1mm]
\dsm &\equiv& m_2^2 - m_1^2
&=& 8.1 \times 10^{-5}\, \mbox{eV}^2.
\end{array}
\ee
Denoting $t_{ij} \equiv \tan{\theta_{ij}}$,
a crucial feature is that $t_{12} < 1$ while $m_1 < m_2$.
We do not know yet whether $m_{1,2} < m_3$ (normal spectrum)
or $m_{1,2} > m_3$ (inverted spectrum).
Let us define
\be
\label{R0}
R \equiv \frac{m_2^2 - m_1^2}{m_3^2 - m_2^2}.
\ee
Then,
\be\label{Rnum}
\left| R \right| = \frac{\dsm}{\dam} \approx 3.7 \times 10^{-2}
\ee
is small,
but the sign of $R$ is unknown:
$R > 0$ corresponds to a normal spectrum,
$R < 0$ to an inverted one.

From $U^T \mathcal{M}_\nu\, U = \mbox{diag} \left( m_1, m_2, m_3 \right)$,
the two defining relations for case C are given by 
\be
\begin{array}{rclcrcl}
\left( \mathcal{M}_\nu \right)_{\mu\mu} &=& 0
& \ \Leftrightarrow \ & 
\left( c_{23} s_{12} + \epsilon s_{23} c_{12} \right)^2 \tilde m_1 + 
\left( c_{23} c_{12} - \epsilon s_{23} s_{12} \right)^2 \tilde m_2 + 
s_{23}^2 c_{13}^2 m_3 &=& 0,
\\
\left( \mathcal{M}_\nu \right)_{\tau\tau} &=& 0
& \ \Leftrightarrow \ & 
\left( s_{23} s_{12} - \epsilon c_{23} c_{12} \right)^2 \tilde m_1 + 
\left( s_{23} c_{12} + \epsilon c_{23} s_{12} \right)^2 \tilde m_2 + 
c_{23}^2 c_{13}^2 m_3 &=& 0,
\end{array}
\label{E}
\ee
where $\epsilon \equiv s_{13} e^{i \delta}$,
$\tilde m_1 \equiv m_1 e^{2 i \rho}$,
and
$\tilde m_2 \equiv m_2 e^{2 i \sigma}$.
This is a system of two linear equations for the two variables
$\tilde m_1 / m_3$ and $\tilde m_2 / m_3$,
which has the solution~\cite{xing1,lavoura}
\be
\label{1,2}
\begin{array}{rcl}
{\displaystyle \frac{\tilde m_1}{m_3}} &=&
{\displaystyle \frac{1}{u}
\left( 1 - \frac{1}{t_{12} z} \right)},
\\*[3mm]
{\displaystyle \frac{\tilde m_2}{m_3}} &=&
{\displaystyle \frac{1}{u} 
\left( 1 + \frac{t_{12}}{z} \right)},
\end{array}
\ee
where
\ba
z &\equiv& \epsilon \tan{2 \theta_{23}},
\label{zoriginal}
\\
u &\equiv& \frac{- 1 + 2 \epsilon \cot{2 \theta_{23}} \cot{2 \theta_{12}}
- \epsilon^2}{c_{13}^2}.
\label{uoriginal}
\ea
Note that,
while $s_{13}$ is known to be small,
$\left| \tan{2 \theta_{23}} \right|$ is possibly large,
because the atmospheric mixing angle $\theta_{23}$
could be close to $45^\circ$. 
Therefore,
$|z| = s_{13} \left| \tan{2 \theta_{23}} \right|$ is totally unknown.
For precisely the same reason---$\left| \epsilon \right|$ is small
while $\left| \tan{2 \theta_{23}} \right|$
is possibly large---the parameter $u$ must be close to $-1$.

It follows from~(\ref{1,2}) that
\be
s_{12}^2\, \frac{\tilde m_1}{m_3} + c_{12}^2\, \frac{\tilde m_2}{m_3}
= \frac{1}{u},
\label{eul}
\ee
which effectively eliminates $z$ from~(\ref{1,2}).
Since $u \simeq -1$,
(\ref{eul}) indicates that an inverted neutrino mass spectrum,
for which both $\left| \tilde m_1 / m_3 \right|$
and $\left| \tilde m_2 / m_3 \right|$ are larger than 1,
is in general compatible with case C.
Conversely,
a normal neutrino mass spectrum can only be compatible with~(\ref{eul})
if $\left| \tilde m_1 / m_3 \right|$
and $\left| \tilde m_2 / m_3 \right|$ are very close to 1.
Thus,
\emph{case C is compatible with a normal neutrino mass spectrum
only if the neutrinos are almost degenerate}.


\paragraph{The special case $s_{13} = \cos{2 \theta_{23}} = 0$:}
The derivation of~(\ref{1,2}) from~(\ref{E})
breaks down when the latter system of equations is singular.
This happens when $s_{13} = 0$,
in which case the system~(\ref{E}) is rewritten as 
\be
\begin{array}{ccccl}
0 & = & 
\left( \mathcal{M}_\nu \right)_{\mu\mu} + 
\left( \mathcal{M}_\nu \right)_{\tau\tau} &=& 
s_{12}^2 \tilde m_1 + c_{12}^2 \tilde m_2 + m_3, 
\\
0 & = & 
\left( \mathcal{M}_\nu \right)_{\mu\mu} - 
\left( \mathcal{M}_\nu \right)_{\tau\tau} &=& 
\left( c_{23}^2 - s_{23}^2 \right)
\left( s_{12}^2 \tilde m_1 + c_{12}^2 \tilde m_2 - m_3 
\right), 
\end{array}
\label{ES}
\ee
whence it follows that $\cos{2 \theta_{23}} = 0$ too.
The system~(\ref{E}) then reduces to only one equation
for the two complex unknowns $\tilde m_1 / m_3$ and $\tilde m_2 / m_3$:
\be
s_{12}^2 \frac{\tilde m_1}{m_3} + c_{12}^2 \frac{\tilde m_2}{m_3} + 1 = 0.
\label{sole}
\ee
It follows from~(\ref{sole}) that $s_{13} = \cos{2 \theta_{23}} = 0$ implies
\emph{an inverted mass spectrum} for the neutrinos.
Indeed,
if both $\left| \tilde m_1 / m_3 \right|$
and $\left| \tilde m_2 / m_3 \right|$ are smaller than 1,
then~(\ref{sole}) cannot hold.

The general solution to~(\ref{sole}) is
\begin{equation}
\label{special}
\frac{\tilde m_1}{m_3} = - 1 + \frac{1}{t_{12} z^\prime},
\quad
\frac{\tilde m_2}{m_3} = - 1 - \frac{t_{12}}{z^\prime},
\end{equation}
where $z^\prime$ is a \emph{free} complex parameter.
Let us compare~(\ref{special}) with~(\ref{1,2}).
Starting from~(\ref{1,2}),
we perform the limit $\left| \epsilon \right| \to 0$,
$\left| \tan{2 \theta_{23}} \right| \to \infty$
while keeping $z = \epsilon \tan{2 \theta_{23}}$ fixed. 
In that limit,
the quantity $u$ of~(\ref{uoriginal}) becomes equal to $-1$
and one obtains~(\ref{special}) with $z^\prime = z$.
Thus,
the special case~(\ref{special})
is obtained as a smooth limit of the general case~(\ref{1,2}),
with an inverted mass spectrum.


\paragraph{The general case:} We define
\ba
\zeta &\equiv& \tan{2 \theta_{12}}\, \mbox{Re}\, z - 1
\no &=& \tan{2 \theta_{12}} \tan{2 \theta_{23}}\, s_{13} \cos{\delta} - 1.
\label{zeta}
\ea
We then derive
\be
\left| u \right|^2 = 1
+ \frac{4 s_{13}^2 \zeta \left( \zeta + c_{13}^2 \right)}
{c_{13}^4 \tan^2{2 \theta_{12}} \left| z \right|^2}.
\label{uz}
\ee
The solution~(\ref{1,2}) allows one to compute
\be
\label{ratio}
\frac{m_2^2 - m_1^2}{m_3^2} = \frac{1}
{\left| u z \right|^2}
\left( \frac{1}{t_{12}^2} - t_{12}^2 \right) \zeta.
\ee
Therefore,
$\zeta$ must be positive.
From~(\ref{ratio}) one further sees that
\be
m_3^2 = \dsm\, \frac{\sin^2{2 \theta_{12}}}{4 \cos{2 \theta_{12}}}\,
\frac{|u|^2 s_{13}^2 \tan^2{2 \theta_{23}}}{\zeta}.
\label{m3}
\ee
The solution~(\ref{1,2}) also allows one to derive a quadratic equation 
\be
\label{quadratic}
a \zeta^2 - b \zeta - c = 0
\ee
for $\zeta$.
Its coefficients are functions of $\theta_{12}$,
$\theta_{13}$,
and $R$:
\ba
a &=& \frac{4 s_{13}^2 R}{c_{13}^4 \tan^2{2 \theta_{12}}},
\no
b &=& \frac{1}{t_{12}^2} - t_{12}^2 +
R \left( 1 - t_{12}^2 \right)
\left[ 1 + \frac{s_{13}^2}{c_{13}^2}
\left( 1 - \frac{1}{t_{12}^2} \right) \right],
\label{coefficients} \\
c &=& R. \nonumber
\ea
Notice that $b$ is positive and of order $1$,
while $a c \propto s_{13}^2 R^2$ is also positive but very small.
Hence,
$\sqrt{b^2 + 4 a c} > b > 0$.
For each sign of $R$,
there is one positive and one negative solution of~(\ref{quadratic}). 
Since $\zeta$ must be positive,
the negative solutions are discarded and we end up with
\be
\label{zeta12}
\zeta_1 = \frac{\sqrt{b^2 + 4 a c} + b}{2 a}
\simeq \frac{b}{a}\ \mbox{for}\ R > 0 
\quad \mbox{and} \quad 
\zeta_2 = \frac{\sqrt{b^2 + 4 a c} - b}{- 2 a}
\simeq - \frac{c}{b}\ \mbox{for}\ R < 0.
\ee
The solution $\zeta_1 \gg 1$ corresponds to a normal spectrum
($R > 0$),
the solution $\zeta_2 \approx 0$ corresponds to an inverted spectrum
($R < 0$).

We shall assume the mixing angles,
the atmospheric mass-squared difference $\dam$,
and the solar mass-squared difference $\dsm$ to be known---see~(\ref{angles})
and~(\ref{m2diff}).
These are five quantities.
On the other hand,
after removing unphysical phases from the mass matrix~(\ref{C}),
for instance by assuming its first row and first column to be real,
we see that that mass matrix contains five physical parameters---four
moduli and one phase.
This suggests that,
from the known mass-squared differences and mixing angles,
we should be able to predict the absolute neutrino mass scale~\cite{xing2}
and also the phases $\delta$,
$\rho$,
and $\sigma$~\cite{FGM,xing1}.
The phase $\delta$,
or at least its cosine,
is determined from $\zeta$ through~(\ref{zeta}).
The absolute mass scale is given by $m_3$ and determined from~(\ref{m3}),
using $|u| \approx 1$.
As far as the phases $\rho$ and $\sigma$ are concerned,
the only observable which is realistically sensitive to them
is the effective mass $\mbb$
in neutrinoless $\beta\beta$ decay~\cite{rodejohann}. 
This effective mass is given by 
\ba
\mbb &=& m_3 \left| c_{13}^2 c_{12}^2 \frac{\tilde m_1}{m_3}
+ c_{13}^2 s_{12}^2 \frac{\tilde m_2}{m_3}
+ \left( \epsilon^\ast \right)^2 \right|
\no &=& m_3 \left| \frac{c_{13}^2}{u} 
\left( 1 - \frac{2}{z \tan 2\theta_{12}} \right) 
+ \left( \epsilon^\ast \right)^2 \right|.
\label{mbb}
\ea
Our strategy may thus be summarized in the following way:
\begin{equation}\label{strategy}
\dam,\ \dsm,\ \theta_{12},\ \theta_{23},\ \theta_{13} 
\ \to \ \delta,\ m_3,\ \mbb.
\end{equation}
%


\paragraph{Normal spectrum:}
In this case,
using~(\ref{zeta12}) we obtain
\begin{equation}\label{delta_n}
\zeta_1 + 1
= \tan{2 \theta_{12}} \tan{2 \theta_{23}}\, s_{13} \cos{\delta}
\simeq \frac{c_{13}^4}{R s_{13}^2 \cos 2\theta_{12}}.
\end{equation}
According to our strategy,
(\ref{delta_n}) determines $\delta$.
This is not very useful,
since that phase will be very difficult to measure;
but a useful inequality following from~(\ref{delta_n}) is 
\begin{equation}\label{ineq1}
\left| \tan{2 \theta_{23}} \right| \gtrsim
\frac{c_{13}^4}{R s_{13}^3 \sin{2 \theta_{12}}}.
\end{equation}
Using the $3\sigma$ bound on $s_{13}$
and the best-fit value of $\theta_{12}$ in~(\ref{angles}),
together with the approximate value of $R$ in~(\ref{Rnum}),
we find $\left| \tan{2 \theta_{23}} \right| \gtrsim 2640$;
for a smaller $s_{13}$
the lower bound on $\left| \tan{2 \theta_{23}} \right|$ is even larger.
Therefore,
\begin{equation}
\theta_{23} = 45^\circ
\end{equation}
for all practical purposes.

In order to determine $m_3$,
we may reformulate $|u|^2$ of~(\ref{uz}) by writing
\begin{equation}
|u|^2 = 1 +
\frac{4 s_{13}^2 \zeta_1 \left( \zeta_1 + c_{13}^2 \right)}
{c_{13}^4 \left( \zeta_1 + 1 \right)^2}\, \cos^2{\delta}
\simeq 
1 + \frac{4 s_{13}^2 \cos^2{\delta}}{c_{13}^4},
\end{equation}
where we have used $\zeta_1 \gg 1$. 
Therefore,
to a good approximation $|u|^2 \simeq 1$ and,
with~(\ref{m3}) and~(\ref{delta_n}),
we obtain 
\begin{equation} \label{m3_n}
m_3 \simeq \frac{\sin{2 \theta_{12}}}{2}\, t_{13}^2
\left| \tan{2 \theta_{23}} \right| \frac{\dsm}{\sqrt{\dam}}.
\end{equation}
According to our strategy,
(\ref{m3_n}) determines the absolute neutrino mass scale.
It is again instructive to convert this into an inequality.
Using~(\ref{ineq1}) we find 
\begin{equation}\label{lower}
m_3 \gtrsim \frac{c_{13}^2 \sqrt{\dam}}{2 s_{13}}.
\end{equation}
With the best-fit value for $\dam$
and the $3 \sigma$ bound on $s_{13}$,
one obtains $m_3\, \gtrsim\, 0.1\, \mathrm{eV}$.
Thus,
for a normal spectrum case C predicts
\emph{quasi-degeneracy of the neutrinos}~\cite{ma}
and \emph{maximal atmospheric neutrino mixing}.
The lower bound on $m_3$
is in such a range that $\sum_j m_j \simeq 3 m_3$
might eventually be extracted
from the large-scale structure of the universe
and the cosmic microwave background~\cite{spergel,barger,hannestad,elgaroy}. 
If $m_3$ is larger than $0.3\, \mathrm{eV}$,
then the neutrino masses lie in the sensitivity range
of the KATRIN experiment~\cite{KATRIN}. 

Turning at last to $\mbb$,
with the methods and approximations used before we find,
from~(\ref{mbb}),
\begin{equation}
\mbb \simeq m_3,
\end{equation}
possibly in the sensitivity range of present,
and certainly in the sensitivity range of future,
neutrinoless $\beta\beta$ decay experiments~\cite{betabeta}.


\paragraph{Inverted spectrum:}
For this spectrum we find from~(\ref{zeta12}) that
\begin{equation}\label{p}
\zeta_2 + 1 = \tan{2 \theta_{12}} \tan{2 \theta_{23}}\, s_{13} \cos{\delta}
\simeq 1 + \frac{\sin^2{2 \theta_{12}} |R|}{4 \cos{2 \theta_{12}}}.
\end{equation}
Since $R$ is small,
$\tan{2 \theta_{12}} \tan{2 \theta_{23}}\, s_{13} \cos{\delta}$
is close to 1~\cite{FGM}. 
From~(\ref{p}) one derives the approximate inequality
\begin{equation}\label{ineq2}
\left| \tan{2 \theta_{23}} \right| s_{13} \gtrsim 
\frac{1}{\tan 2\theta_{12}},
\end{equation}
which may be useful in the future for testing case C. 

We turn to the absolute neutrino mass scale~\cite{xing2}.
Because $\zeta_2 \sim |R|$ is small,
and because of the inequality~(\ref{ineq2}),
$|u|$ as computed from~(\ref{uz}) is,
to a good approximation,
equal to 1.
Then~(\ref{m3}) leads to
\begin{equation}
m_3 \simeq \sqrt{\dam}\, s_{13} \left| \tan{2 \theta_{23}} \right|.
\label{m33}
\end{equation}
From $\dsm \ll \dam$ and from the definition of $\dam$ we then have
\begin{equation}
m_1 \simeq m_2 \simeq \sqrt{\left( 1 + 
s_{13}^2 \tan^2{2 \theta_{23}} \right) \dam} 
\end{equation}
and 
\begin{equation}\label{sum}
\sum_{j=1}^3 m_j \simeq 
\left( 2\, \sqrt{1 + s_{13}^2 \tan^2{2 \theta_{23}}} + 
s_{13} \left| \tan{2 \theta_{23}} \right| \right) 
\sqrt{\dam}.
\end{equation}
An important feature of the inverted spectrum
is that $s_{13} \left|\tan{2 \theta_{23}} \right|$
is of order 1 or larger,
cf.~(\ref{ineq2}).
Therefore,
just as in the case of the normal spectrum,
the observable~(\ref{sum})
has an order of magnitude interesting for cosmology.

Using~(\ref{mbb}),
with $|u| \simeq 1$ and neglecting $\left( \epsilon^\ast \right)^2$,
we find
\begin{equation}
\mbb \simeq  
m_3 \left| 1 - 
\frac{2}{z \tan{2 \theta_{12}}} \right| 
= m_3 
\left[ 1 - \frac{4 \zeta_2}
{\left( s_{13} \tan{2 \theta_{23}} \tan{2 \theta_{12}} \right)^2}
\right]^{1/2} 
\simeq m_3.
\end{equation}
With~(\ref{m33}) this produces the remarkable relation
\begin{equation}\label{relation}
\mbb \simeq \sqrt{\dam}\, s_{13} \left| \tan{2 \theta_{23}} \right|,
\end{equation}
which may in the future allow
an experimental test of case C
with an inverted spectrum.


\section{Conclusions} \label{concl}

In this paper we have shown that
all seven viable neutrino mass matrices with two texture zeros
which were found in \cite{FGM} can be embedded in models
with at most three scalar $SU(2)_\mathrm{L}$ triplets,
without adding new fermions or Higgs doublets to the SM.
Utilizing one of the methods of \cite{GJLT},
we have used horizontal groups of the type $\mathbbm{Z}_n$.
The model for the mass matrix~(\ref{C}),
with three scalar triplets and a symmetry $\mathbbm{Z}_4$,
was discussed in detail in Section~\ref{model},
whereas the symmetry realizations
of the other six cases were studied in the appendix.
To make the triplet VEVs small and the new scalars heavy,
we have used the type-II seesaw mechanism~\cite{ma1};
the heavy triplets are integrated out at about $10^{13}\, \mathrm{GeV}$.

It is interesting to note that texture zeros
in the neutrino mass matrix are stable
under the renormalization group evolution of $\mathcal{M}_\nu$,
as long as there is only one (light) Higgs
doublet~\cite{lavoura}, 
which is the case in the models that we have produced. 
The reason is that in this case the evolution of the mass matrix
is described as $\mathcal{M}_\nu (\mu)
= I \left( \mu, \mu_0 \right) \mathcal{M}_\nu \left( \mu_0 \right)
I \left( \mu, \mu_0 \right)$,
where $I \left( \mu, \mu_0 \right)$ is a diagonal and positive
$3 \times 3$ matrix~\cite{chankowski}.
If an element of $\mathcal{M}_\nu$ is zero
at the renormalization scale $\mu_0$,
then it remains so for all other renormalization scales $\mu$.

We have shown that the mass matrix~(\ref{C}),
christened by~\cite{FGM} `case C',
leads to an interesting phenomenology,
possibly testable in the near future. 
It allows for both types of neutrino mass spectra,
either normal or inverted.
For both spectral types the effective mass $\mbb$ 
in neutrinoless $\beta\beta$ decay is approximately equal to $m_3$. 

The case with a normal spectrum is tightly constrained: 
the neutrinos must be quasi-degenerate,
with a mass larger than about $0.1\, \mathrm{eV}$,
and the atmospheric mixing angle is equal to $45^\circ$
for all practical purposes.
The lower bound on the neutrino masses scales with the inverse of $s_{13}$,
i.e.\ a smaller upper bound on $s_{13}$ will require larger neutrino masses. 
Recently,
cosmology has become important
in constraining the sum of all light-neutrino masses;
current results set an upper limit 
$\sum_j m_j \simeq 3 m_3 \lesssim 1\,
\mathrm{eV}$~\cite{spergel,barger,hannestad,elgaroy}.  
With~(\ref{lower}) and the best-fit value of $\dam$,
this upper limit translates into $s_{13}^2 \gtrsim 0.005$.
The cosmological bound may be improved by one order of magnitude
in the near future~\cite{elgaroy}; 
lowering the bound on $\sum_j m_j$
will imply a rise of the lower bound on $s_{13}^2$.
Thus,
future measurements of $s_{13}^2$~\cite{huber},
$\sum_j m_j$,
and $\mbb$,
together with the bound~(\ref{lower}) and $\sum_j m_j/\mbb \simeq 3$,
will test case C with the normal spectrum.

The case with an inverted spectrum is characterized
by the product $\sin \theta_{13} \left| \tan{2 \theta_{23}} \right|$
being of order one or larger; 
thus,
a tighter upper bound on $s_{13}$ will imply
a $\theta_{23}$ closer to $45^\circ$. 
The value of that product also determines
the absolute neutrino mass scale. 
Though the neutrino masses in the inverted-spectrum case
do not need to be as large as those of the normal spectrum,
they still present a very good discovery potential,
cf.\ for instance $\mbb \simeq m_3 \gtrsim 0.02\,
\mathrm{eV}$.\footnote{This lower bound was calculated
by using~(\ref{ineq2}) and~(\ref{m33})
together with the best-fit values for $\dam$ and $\theta_{12}$.}

In any case,
there are good prospects that the predictions of case C
are either confirmed or disproved in the near future.


\begin{appendix}

\setcounter{equation}{0}
\renewcommand{\theequation}{A\arabic{equation}}

\section{Realization of the other two-texture-zero cases \\
through a symmetry}

In Section~\ref{model} we dealt with the realization
of the case C of~\cite{FGM} in a triplet model. 
Here we discuss the other six viable two-texture-zero cases of~\cite{FGM},
defined as
\begin{eqnarray}
&&
\mbox{case A$_1$:} \quad \mathcal{M}_\nu \sim
\left( \begin{array}{ccc} 
0 & 0 & \times \\ 0 & \times & \times \\ \times & \times & \times 
\end{array} \right), \quad
\mbox{case A$_2$:} \quad \mathcal{M}_\nu \sim 
\left( \begin{array}{ccc} 
0 & \times & 0 \\ \times & \times & \times \\ 0 & \times & \times 
\end{array} \right),
\\
&&
\mbox{case B$_1$:} \quad \mathcal{M}_\nu \sim 
\left( \begin{array}{ccc} 
\times & \times & 0 \\ \times & 0 & \times \\ 0 & \times & \times 
\end{array} \right), \quad
\mbox{case B$_2$:} \quad \mathcal{M}_\nu \sim 
\left( \begin{array}{ccc} 
\times & 0 & \times \\ 0 & \times & \times \\ \times & \times & 0
\end{array} \right),
\\
&&
\mbox{case B$_3$:} \quad \mathcal{M}_\nu \sim
\left( \begin{array}{ccc} 
\times & 0 & \times \\ 0 & 0 & \times \\ \times & \times & \times 
\end{array} \right), \quad
\mbox{case B$_4$:} \quad \mathcal{M}_\nu \sim
\left( \begin{array}{ccc} 
\times & \times & 0 \\ \times & \times & \times \\ 0 & \times & 0
\end{array} \right).
\end{eqnarray}
\paragraph{Cases B$_1$ and B$_2$:}
With $\omega = \exp{\left( 2 i \pi / 3 \right)}$,
we fix $p_e = 1$, $p_\mu = \omega$, and $p_\tau = \omega^2$,
i.e.\ we assume a horizontal symmetry $\mathbbm{Z}_3$.
The bilinears
$D_{\alpha \mathrm{L}}^\mathrm{T} C^{-1} D_{\beta \mathrm{L}}$
transform according to
\begin{equation}
\left( \begin{array}{ccc}
1 & \omega & \omega^2 \\ \omega & \omega^2 & 1 \\ \omega^2 & 1 & \omega
\end{array} \right).
\end{equation}
Two scalar triplets are then sufficient
to realize textures B$_1$ and B$_2$ as models:
\begin{eqnarray}
\mathrm{B}_1: && \Delta_1 \to \Delta_1, \quad \Delta_2 \to \omega^2
\Delta_2; \\
\mathrm{B}_2: && \Delta_1 \to \Delta_1, \quad \Delta_2 \to \omega
\Delta_2.
\end{eqnarray}

\paragraph{Cases A$_{1,2}$ and B$_{3,4}$:}
We firstly make the following observation: 
two non-zero entries in the same column (or row)
of $\mathcal{M}_\nu$
cannot originate in Yukawa couplings to the same scalar triplet.
Indeed,
if that were the case then we would have $p_\alpha = p_\beta$
for two flavours $\alpha$ and $\beta$.
The Yukawa couplings to the relevant scalar triplet
would then provide \emph{three} non-zero entries in $\mathcal{M}_\nu$,
at the positions $\left( \alpha, \alpha \right)$,
$\left( \alpha, \beta \right) = \left( \beta, \alpha \right)$,
and $\left( \beta, \beta \right)$.
Only cases A$_1$ and A$_2$ display such a non-zero ``square,''
but one immediately sees that with 
$p_\gamma \neq p_\alpha = p_\beta$ one cannot generate those two textures.

Secondly,
we note that the textures A$_{1,2}$ and B$_{3,4}$
(and C) all have one column (and row)
of non-zero matrix elements in $\mathcal{M}_\nu$.
Following the previous paragraph,
those three matrix elements must be generated
by the Yukawa couplings to three \emph{different} scalar triplets.
Thus,
except for B$_1$ and B$_2$,
all other textures of~\cite{FGM} require at least three
$SU(2)_\mathrm{L}$ triplets to be realized within a model
featuring an Abelian  horizontal symmetry.
In the following we will see that three scalar
triplets are indeed sufficient for textures A$_{1,2}$ and B$_{3,4}$;
for texture C we had explicitly demonstrated the same
in Section~\ref{model},
by using a symmetry $\mathbbm{Z}_4$.

One can check that the textures A$_{1,2}$ and B$_{3,4}$
cannot be realized through a symmetry $\mathbbm{Z}_4$.
On the other hand,
it is possible to use phase factors $1$,
$-1$,
and $p$,
provided $p^2 \neq \pm 1$.
Let us take $p_\mu = 1$ and $p_\tau = -1$,
while leaving $p_e$ free for the moment.
Since $p_e^2$ must be different from $\pm 1$,
we can take for instance $p_e = \exp{\left( 2 i \pi / 3 \right)}$; 
this,
together with $p_\tau = -1$, 
means that the horizontal symmetry is $\mathbbm{Z}_6$.
The bilinears
$D_{\alpha \mathrm{L}}^\mathrm{T} C^{-1} D_{\beta \mathrm{L}}$
transform as 
\begin{equation}
\left( \begin{array}{ccc}
p_e^2 & p_e & - p_e \\ p_e & 1 & - 1 \\ - p_e & - 1 & 1
\end{array} \right),
\end{equation}
and we realize textures A$_1$ and A$_2$
with three scalar triplets transforming as 
\begin{eqnarray}
\mathrm{A}_1: & & \Delta_1 \to \Delta_1, \quad
\Delta_2 \to - \Delta_2, \quad
\Delta_3 \to - p_e^\ast \Delta_3; \\
\mathrm{A}_2: & & \Delta_1 \to \Delta_1, \quad
\Delta_2 \to - \Delta_2, \quad
\Delta_3 \to p_e^\ast \Delta_3.
\end{eqnarray}
An analogous method may be employed to enforce textures B$_{3,4}$.

All these realizations of the two-texture-zero cases of~\cite{FGM}
require only a single Higgs doublet.
One of the scalar triplets
(in our notation,
$\Delta_1$)
can always be assumed to transform trivially under the horizontal symmetry.
Therefore,
the discussion of the type-II seesaw mechanism~\cite{ma1}
at the end of Section~\ref{model} applies to all these realizations.

\end{appendix}


\vskip 10mm
\noindent \textbf{Acknowledgement}:
The work of L.L. was supported by the Portuguese
\textit{Fun\-da\-\c c\~ao para a Ci\^encia e a Tecnologia}
under the project U777--Plurianual.


\end{document}